# A bibliometric analysis of research based on the Roy Adaptation Model: a contribution to Nursing

## Paulina Hurtado-Arenas[1] and Miguel R. Guevara[2]

[1]  Universidad de Valparaíso; paulina.hurtado@uv.cl

[2]  Universidad de Playa Ancha; miguel.guevara@upla.cl

## Abstract

**Objective.** To perform a modern bibliometric analysis of the research based on the Roy Adaptation Model, a founding nursing model proposed by Sor Callista Roy in the1970s.

**Method.** A descriptive and longitudinal study. We used information from the two dominant scientific databases, Web Of Science and SCOPUS. We obtained 137 publications from the Core Collection of WoS, and 338 publications from SCOPUS. We conducted our analysis using the software Bibliometrix, an R-package specialized in creating bibliometric analyses from a perspective of descriptive statistics and network analysis, including co-citation, co-keyword occurrence and collaboration networks.

**Results.** Our quantitative results show the main actors around the research based on the model and the founding literature or references on which this research was based. We analyze the main keywords and how they are linked. Furthermore, we present the most prolific authors both in number of publications and in centrality in the network of coauthors. We present the most central institutions in the global network of collaboration.

**Conclusions.** We highlight the relevance of this theoretical model in nursing and detail its evolution. The United States is the dominant country in production of documents on the topic, and the University of Massachusetts Boston and Boston College are the most influential institutions. The network of collaboration also describes clusters in Mexico, Turkey and Spain. Our findings are useful to acquire a general vision of the field.

Keywords

*Roy adaptation model; bibliometrix; bibliometrics; nursing; co-authorship network; centrality*

## 1 Introduction

Nursing as a discipline goes beyond the medical context. It acts not only in illness situations but also in the care of people, considering different aspects such as the social context, dignity or well-being. Nursing is centered on the care of the person that interacts with his or her environment. This idea includes not only defined care tasks but also a focalization in support, self-respect or advice (Raile & Marriner, 2010). In this line, in the sixties and seventies, several theoretical frameworks were proposed to substantiate the idea of Nursing as a discipline distinct or not peripheral to the medical process. Some Nursing Models prevent the practice of Nursing from being based only on routines, rituals or intuitions. One of those models is the Roy Adaptation Model that has had a broad diffusion around the world over time.

The Roy Adaptation Model (RAM) was proposed by Sor Callista Roy in the seventies and in further updates (Callista Roy & Anderson, 1999; C. Roy, 2011a; Sister Callista Roy, 1984). The model is a middle-range theoretical framework that helps nursing practice to conduct patients' care in a more humane and even philosophical manner. Roy proposes to promote adaptation in four aspects or adaptive models that contribute to quality of life, health and dying with dignity (C. Roy, 2011a). The adaptation of individuals and groups is at the core of a transformation of society in one that promotes dignity.

Because of this deep conception of nursing practice and a solid kit of instruments available to conduct further research (Barone, Roy, & Frederickson, 2008), the Roy Adaptation Model has been positioned as one of the most implemented models over decades, not only in the United States but also around the world.  In this line, Sor Callista Roy has reviewed in past years the different paths that her theory has navigated by analyzing the research based on the model (C. Roy, 2011b; Callista Roy, 2013). These efforts by Sor Callista and colleagues have been undertaken to distinguish the different areas where the model has been occupied or the middle theories that have evolved within the framework theory.

Previous analyses are useful to better understand the evolution and application of the main concepts, and the application of the instruments. However, those analyses are not able to process the complete documentation available on the theme. Here is when automatic analysis and quantitative methods are useful to analyze the full spectrum of authors, publications, institutions and other important actors surrounding the Adaptation Model.

Analyzing quantitatively (and automatically) the whole bibliography on a theme, and applying statistical methods, is known as a bibliometric analysis. This kind of analysis helps to generate a 'big picture' of the topic analyzed, in this case the Roy's Adaptation Model. A bibliometric analysis is composed of different types of analysis that can be computed based on information within a paper and among papers, such as the words of the text, the information about authors, and the references and citations gathered by the publication (Sengupta, 2009). The main analyses are usually: descriptive statistics, analysis of authors, analysis of keyword, analysis of references, and a set of network analyses such as co-citation, co-authorship and collaboration.

Bibliometric analyses are not new. They have been created since the seventies (Small, 1973), but are currently more frequent since specialized software has been developed in recent years to perform this task automatically (Aria & Cuccurullo, 2017). This is an important point given that most of the calculations to conduct a bibliometric analysis are not easy to compute by manual means or with general purpose software, such as Excel or SPSS. This increasing interest in bibliometric analysis has also been catalyzed for facilities to export bibliographic data granted by the main scientific databases such as Web of Science or SCOPUS, who index specialized literature in the different areas of knowledge.

Bibliometric analyses are rich in the application of data visualization. This is interesting to note, since the founding mother of modern Nursing, Florence Nightingale, was also one of the precursors of applied statistics and the data visualization field with her –now so defunded– *coxcomb* or 'Nightingale Rose', which is a polar graph that showed, in the XIX century, that the mortality of soldiers of the British Army was mainly due to foreseeable causes (related to sanity), rather than to injuries of war (Álvarez, Guevara, & Orellana, 2018; McDonald, 2014). In this study we return to the rich tradition of presenting statistics in a graph or data visualization by presenting a bibliometric analysis of the research based on the Roy's Adaptation Model.

## 2 Literature review

Bibliometric analyses applied to the area of Nursing have been published in the past years with more or less detail in the type of tools or indicators used for the analysis. A review of the history of bibliometric indicators applied to Nursing, and how they are relevant for research in the discipline or for measuring the performance of journals and scholars, was published by Smith and Hazelton (2008) a decade ago.

Studies applied to specific topics in Nursing have been also published, as in the case of the bibliometric analysis conducted by Campbell and collaborators to compare the performance of research in Adult Social Care (Campbell et al., 2015). The authors retrieve roughly two hundred thousand publications from SCOPUS between 1990 and 2011. With this dataset, they computed mainly descriptive statistics of the countries and institutions involved in the research of this topic, analyzing also the collaboration network among UK institutions. A similar, country-oriented study was published by Spanish scholars who analyzed the bibliometric characteristics of papers published in four nursing journals of Spain from 1985 to 1994 (Pardo, Reolid, Delicado, Mallebrera, & García-Meseguer, 2001). Descriptive statistics and statistical tests were run to analyze the differences between the citation performances of each journal, and to characterize the Spanish nursing research.

In terms of Region, Mendoza-Parra and colleagues have analyzed the Latin American Nursing production applying bibliometric indicators. The researchers conducted a deep search on Latin American journals of Nursing, and queried bibliographic information of 5 different sources in the time period between 1959 and 2005. They conclude that nursing research in Latin America is a growing tendency, but that an effort in the organization of the process of publication is still required.

Other studies focus on analyzing terms and their bibliometric characteristics. For example, Anderson and collaborators (Anderson, Keenan, & Jones, 2009) analyzed the differences between five different sets of terms used to represent nursing diagnoses, outcomes, and intervention. They mined more than one thousand references included in the Cumulative Index to Nursing and Allied Health Literature (CINAHL) database in the period between 1982 and 2006. Descriptive statistics and differences between co-author networks are analyzed to conclude the main differences between these sets of terms.

The available resources for conducting bibliometrics analysis in Nursing was studied by Alfonzo and collaborators (Alfonzo, Sakaraida, & Hasting-Tolsma, 2014). The authors summarize the considerations to analyze nursing research to create bibliometric mapping. They propose a step-by-step process in order to successfully create the analysis. They also detail software tools specialized for the task of gathering and processing bibliographic data, and finally they create a bibliometric analysis using the Roy Adaptation Model as a case study. To this end, they queried ProQuest Dissertation Abstracts International database, gathering 111 references. The cited study was limited to The United States and dissertations or thesis. They used the software VOSViewer (van Eck & Waltman, 2009) to generate the report. This software allows the bibliometric analysis with emphasis on indicators based on network analysis, for example, keyword co-occurrence, co-authorship or co-citation. Our analysis is close to the previous study but different in several aspects: first, in the characteristics of the data source, since we use data that is global and based on WoS and SCOPUS; second, in the type of analysis, as we extend the analysis to a more complete set of indicators and visualizations; and third, in the software tool used to generate the report, because we use the recently published software bibliometrix (Aria & Cuccurullo, 2017), which allows other types of analysis if compared with VosViewer.

# 3 Methodology

### 3.1 Data source

In July 2018, we analyzed scientific databases WoS and SCOPUS to find the coverage of the Roy Adaptation Model in each one. Applying an advanced search in both databases, we retrieved 137 publications in WoS Core Collection and 338 publications in SCOPUS.

In the WoS advanced search we used the following query:

*(TS= (roy adaptation model) OR TS=(modelo de adaptación de roy) OR TS=(modelo de adaptacion de roy))*

In the SCOPUS advanced search we used the following query:

*( TITLE-ABS-KEY ( "roy adaptation model" ) ) OR ( TITLE-ABS-KEY ( "modelo de adaptación de roy" ) ) OR ( TITLE-ABS-KEY ( "roy's adaptation model" ) ) OR ( TITLE-ABS-KEY ( "modelo de adaptacion de roy" ) ) )*

### 3.2 Data cleaning

In both databases, we manually reviewed each publication and discarded publications that were not related to the Roy Adaptation Model. We also manually modified fields that could be pointing to the same person but that were written differently, for example, "S.C. Roy" and "C. Roy" were merged.

The databases were exported in .bib files using the functionalities provided by both systems. The dumped files were imported in the specialized software Bibliometrix for further analysis.

We conducted our analysis based (separately) on both datasets. Given the similarities between datasets, in further analysis we used in some cases SCOPUS and in other cases WoS, with the aim of not duplicating analysis and presenting the most relevant findings of each dataset.

### 3.3 Software tools

The analysis was done using the software Bibliometrix (Aria & Cuccurullo, 2017). It is an R-package recently released that surpasses other specialized packages such as iGraph (Csardi & Nepusz, 2006) or GGPlot (Wickham, 2016). It allows the importation of bibliographic data files and specific analysis using descriptive statistics and network analysis.

For general descriptive statistics, we also used the online reports that both databases allow in their respective web sites.

# 4 Results

## 3.1 Descriptive statistics

The first bibliometric indicators are descriptive measures of productivity and performance for producers of science, namely, countries, institutions or scholars. These indicators are provided for the analysis systems of WoS or SCOPUS, and are computed in detail with the software Bibliometrix. Table 1 presents the detail of relevant documents for Roy Adaptation Model in both databases. We observe the wider coverage of SCOPUS that is roughly 2.5 times the production indexed by WoS. In terms of performance, each document has an average of roughly 7 cites, while the H-index according to SCOPUS is 23 and for WoS it is 15. This implies, for example, that the 15 most cited papers in WoS gained at least 15 cites. The statistics related to authors indicate that this field of research on average includes 1.8 authors per document, and that solo-papers are close to 15% of the total documents in both databases. Documents are developed in collaboration between on average 3 institutions.

**Table 1**: Descriptive statistics for bibliography in SCOPUS and WOS

| Measure | SCOPUS | WOS |
|---|---|---|
| Number of Documents | 338 | 137 |
| Sources (Journals, Books, etc.) | 144 | 53 |
| Keywords Database | 1400 | 329 |
| Authors' Keywords (DE) | 500 | 318 |
| Period | 1976 - 2018 | 1977 - 2018 |
| Average citations per documents | 7.746 | 6.117 |
| H-Index | 23 | 15 |
| Authors | 620 | 256 |
| Authors of single-authored documents | 92 | 30 |
| Authors of multi-authored documents | 528 | 226 |
| Authors per Document | 1.83 | 1.87 |
| Collaboration Index | 2.72 | 2.9 |

In Figure 1, we present the detailed annual evolution of production of documents and average total citations per year reported by SCOPUS. We notice that the production of new research based on the Adaptation Model is not constant, but it describes an increasing tendency.

**Figure 1:** Production of documents and average total citations per year.

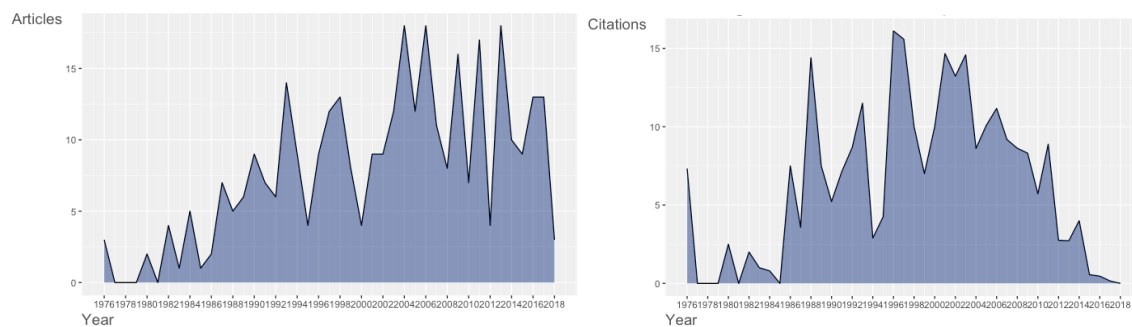

Analyzing the journals where the scientific research of the topic has been published (See Figure 2), we find that the preferred Nursing Journal to publish research related to the Roy Adaptation Model is the Nursing Science Quarterly, followed by the Journal of Advanced Nursing.

**Figure 2:** Production of documents and average total citations per year.

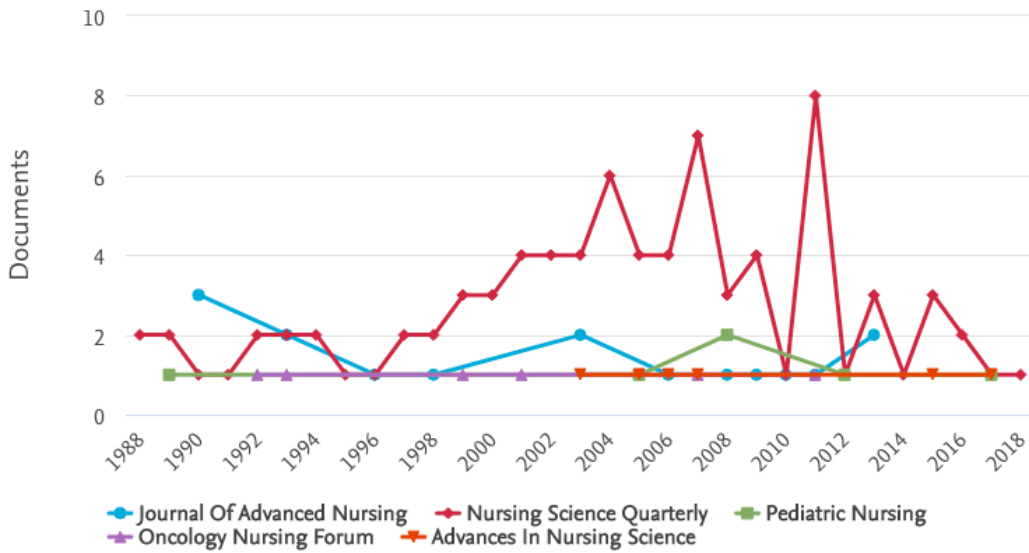

## 3.2 Areas and types of documents

Figure 3 presents the knowledge area where papers have been published and the type of each document according to SCOPUS. As expected, 65% of the documents are indexed in the area of Nursing and 25.9% in the area of Medicine. Also, 86.1% of the papers correspond to articles and roughly 10% to reviews.

**Figure 3:** Knowledge area assigned by SCOPUS to each document and type of documents.

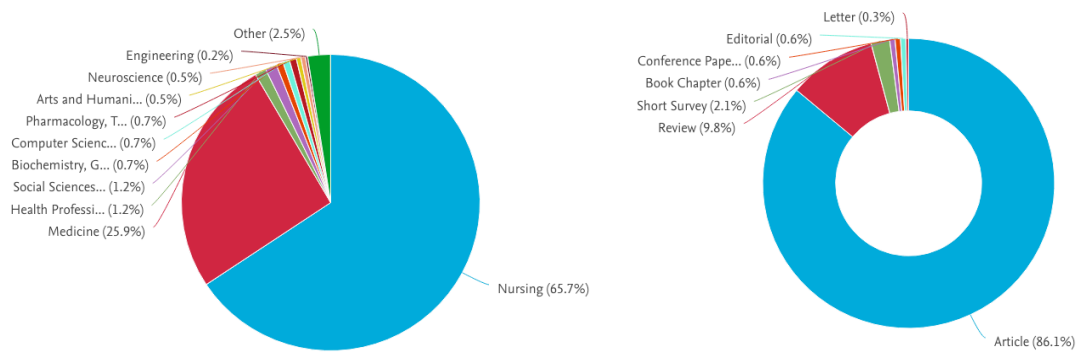

**Source:** SCOPUS

## 3.2 Producers

In terms of producers of documents about the topic (See Figure 4), at country level, the United States is definitely the most productive. Asia, Turkey, Iran, Taiwan, Korea and Japan are ranked in the top ten. Latin America, Brazil and Colombia are also part of the list. Colombia and Taiwan present an important level of collaboration based on the number of publications with other countries (Multiple Country Publications) but their number of publications are limited. This global dissemination could also be linked to the extended coverage of the Roy Adaptation Association (RAA), which has open chapters in Colombia, Japan and Panama.

At institution level, the University of Massachusetts Boston, Boston College and the City University of New York are the highest contributing institutions to the topic according to SCOPUS. The Universidade Federal do Ceara is the only Latin American University included in the top ten contributors (see Figure 4).

**Figure 4:** Top ten countries and Institutions that published documents about the Roy Adaptation Model. (SCP: Single Publication Country. MCP: Multiple Publication Country). Data based on SCOPUS.

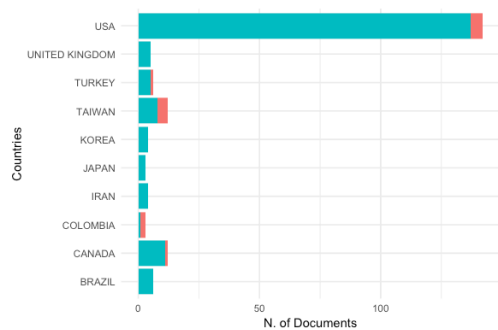
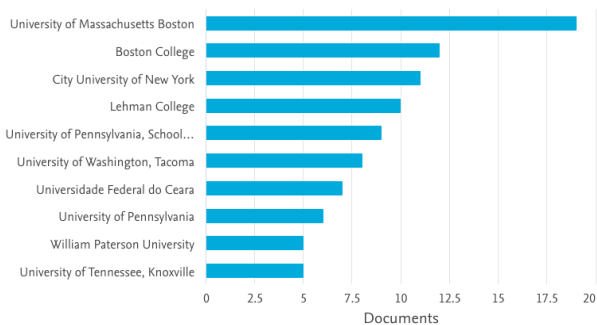

At an individual level, we measure the most productive authors in terms of the number of papers and also in terms of fractional authorship, for example, a paper including two authors counts 0.5 for each one (See Table 2). The latter measure champions solo-authors and reduces the impact of hyperauthorship. The most productive author on the topic is Dr. Jacqueline Fawcett, faculty of the University of Massachusetts Boston. Dr. Fawcett conducted a Roy Adaptation Model-based program of research in that university.

**Table 2**: Ranking of most productive authors in Roy Adaptation Model research based on number of articles published and on fractionalized authorship. Data based on SCOPUS.

| Ranking | Authors | Articles | Ranking | Authors | Fractionalized |
|---|---|---|---|---|---|
| 1 | FAWCETT J | 26 | 1 | FAWCETT J | 10.89 |
| 2 | FREDERICKSON K | 10 | 2 | DOBRATZ MC | 8.5 |
| 3 | DOBRATZ MC | 9 | 3 | ROY C | 7.03 |
| 4 | ROY C | 9 | 4 | FREDERICKSON K | 5.28 |
| 5 | TULMAN L | 6 | 5 | HANNA DR | 3.5 |
| 6 | BUCKNER EB | 5 | 6 | DESANTO-MADEYA S | 3.2 |
| 7 | DESANTO-MADEYA S | 5 | 7 | NEWMAN DM | 3 |
| 8 | SAMAREL N | 5 | 8 | YEH C-H | 3 |
| 9 | ABER C | 4 | 9 | DUNN KS | 2.33 |
| 10 | FOOTE A | 4 | 10 | GAGLIARDI BA | 2.33 |

## 3.2 The intellectual infrastructure of the field

As we analyzed in previous sections, scientific databases include mainly articles. These papers build on previous knowledge developed that was communicated mainly in books, which is the case of the Roy Adaptation Model. To track the intellectual infrastructure of the field, we can find the references most cited by our database of documents. According to WoS, the most cited references are the books about the Roy Adaptation Model in their different editions followed by research related to the theory, reviews, the instruments of the model, and the application of the model to life closure.

**Table 3**: The intellectual infrastructure of the research based on the Roy Adaptation Model. Data based on WoS.

| Reference | Year | Title | Type | Cites |
|---|---|---|---|---|
| **(Callista Roy & Anderson, 1999)** | 1999 | The Roy Adaptation Model | Book | 75 |
| **(C. Roy, 2009)** | 2009 | The Roy Adaptation Model | Book | 43 |
| **(Sister Callista Roy, 1984)** | 1984 | Introduction To Nursing: An Adaptation Model | Book | 15 |
| **(S.C. Roy, 1988)** | 1988 | An explication of the philosophical assumptions of the Roy adaptation model. | Journal | 14 |
| **(C. Roy & Andrews, 1991)** | 1991 | The Roy Adaptation Model: The Definitive Statement | Book | 12 |

| (C. Roy, 1997) | 1997 | Future of the Roy model: Challenge to redefine adaptation | Journal | 11 |
| (Fawcett, 2005) | 2005 | The Roy Adaptation Model. In Contemporary Nursing Knowledge Analysis and Evaluation of Nursing Models and Theories | Book Section | 9 |
| (Perrett, 2007) | 2007 | Review of Roy adaptation model-based qualitative research | Journal | 9 |
| (Barone et al., 2008) | 2008 | Instruments used in roy adaptation model-based research: Review, critique, and future directions | Journal | 7 |
| (Marjorie C. Dobratz, 2002) | 2002 | The Pattern of the Becoming-Self in Death and Dying | Journal | 7 |
| (M.C. Dobratz, 2004) | 2004 | Life-closing spirituality and the philosophic assumptions of the roy adaptation model | Journal | 7 |

To advance this analysis, we map a network (a structure composed of nodes and links) that represents which papers are cited together more frequently. This network, known as co-citation network (Small, 1973), allows analyzing how the main intellectual stream of the topic is related (See Figure 5). The links represent the number of papers that reference a pair of nodes (documents) simultaneously. The wider the link, the bigger the number of papers that include in their references a given pair of nodes. For the case of the nodes, the bigger the size of the node, the bigger the centrality of the document. Centrality in this case is measured for the number of links gathered by the node. This type of centrality is also called *Degree Centrality*.

In the case of the Roy Model, the seminal books on the area, and some papers, provide the Intellectual infrastructure. We can also analyze the links of the central nodes to other peripheral nodes. For example, there is a link to qualitative research. This link is fed for documents citing Roy's book together with Perrett's paper. Similarly, the derivation of a philosophical approach from Roy research is due to articles that increase the links between Roy's books and Dobratz's papers. Other links relate the central literature to Hanna's or Lazarus' documents.

**Figure 5:** Co-Citation Network. Nodes represent documents (books or papers) identified by the first author and the year. Links are given for documents citing both nodes at the same time. Color is assigned according to an algorithm of community detection. Data based on WoS.

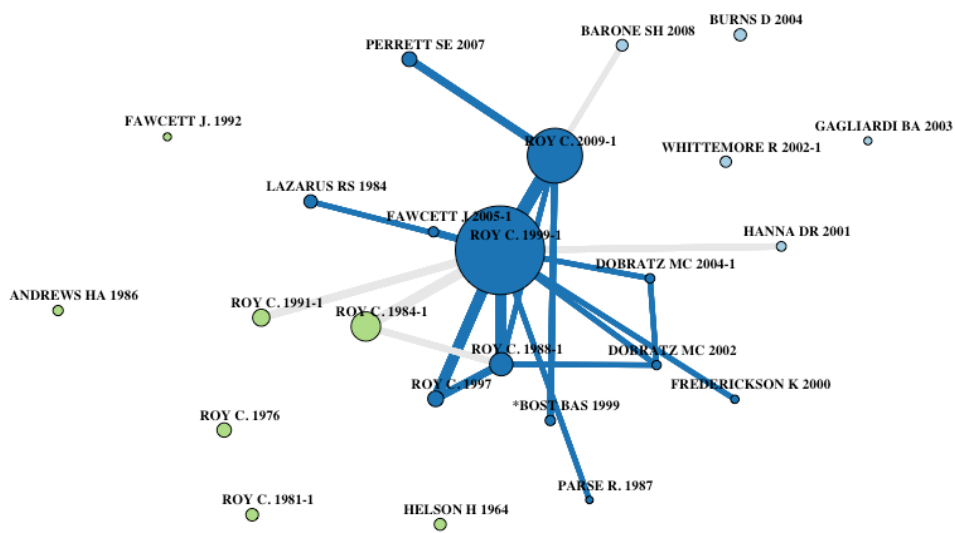

### 3.3 Keyword analysis

Regarding the content of the documents and how they are associated, we can analyze the keywords that authors use to describe their contribution. We analyze this information by mapping similarities between key concepts. For this purpose we map a Keyword co-occurrence network. This network is built with a similar idea to the previous one. In this sense, we draw a link between two keywords if two documents (or more) include both concepts in their descriptors. This network allows us to detect communities of concepts that describe the main topics of the research based on the Roy Model. See colors assigned to nodes in Figure 6.

We observe the main cluster in blue; it corresponds to the documents linking the keyword Health with Adjustment, Cancer or Spirituality. The second cluster of documents (pink), links Social Support to Quality of Life, Women and Breast-Cancer. A third main group (sky blue) of research topics links the word Management to Therapy or Surgery. Other clusters of documents are defined by the keywords Coping Strategies (purple), Scale (light orange), Care (orange) and Mortality (lavender).

**Figure 6:** Keyword co-occurrence Network. Nodes represent keywords. Links are given for documents describing their content with both keywords. Color is assigned according to an algorithm of community detection. Data based on WoS.

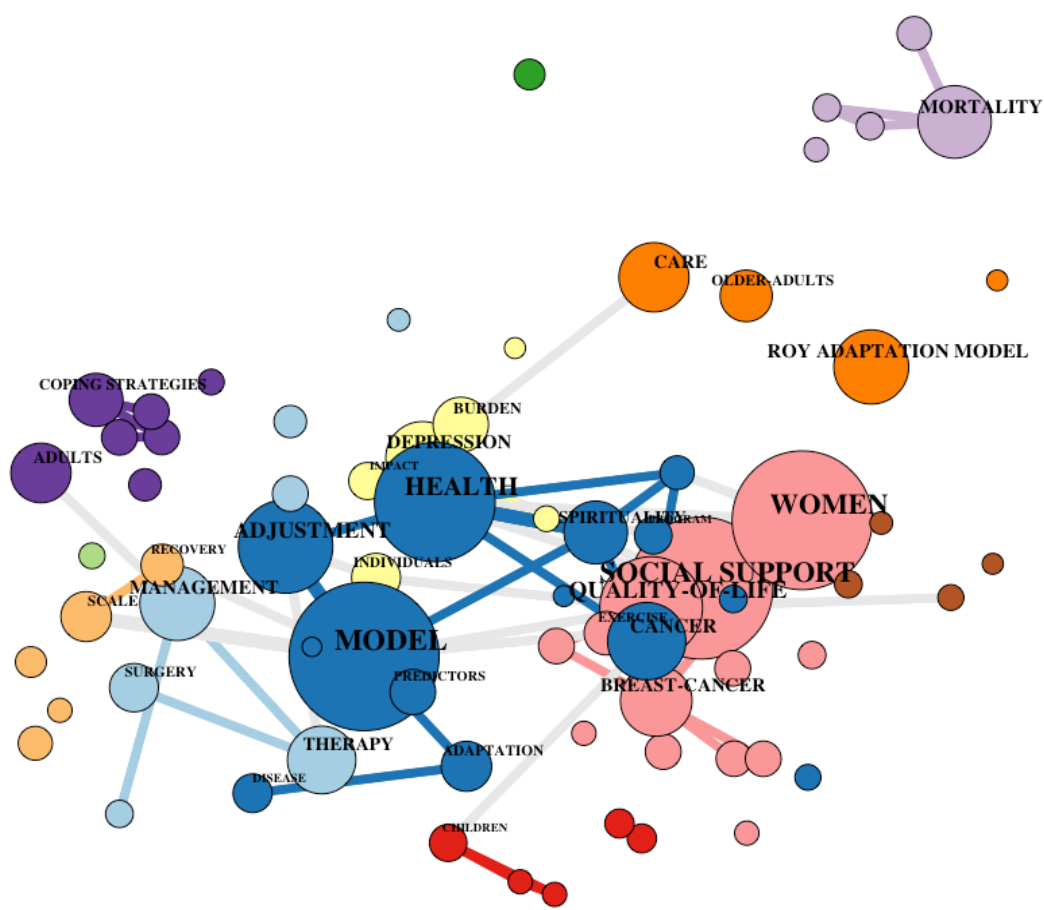

### 3.3 Collaboration

Collaboration in a bibliometric analysis is proxied by co-authorship. The granularity of collaboration can be obtained at an individual, institution or country level, according to the affiliation that authors sign in a document. Here we analyze individual and institutional collaboration around the Roy Model.

Regarding individual collaboration, we map a co-authorship network for each database. For illustrative purposes, in Figure 7 we present the main structure of the WoS network (256 nodes), while in the Supplementary Material we attach the full co-authorship network based on SCOPUS (620 nodes). In this network each node represents an author, and a link is weighted for the number of papers co-authored by two authors. In Figure 7 we note 6 main clusters of authors. The cluster with most authors is composed of the research group of Fawcett, who is also an important hub (a node with high centrality) of the network because she is also connected with other scholars outside her cluster, including Roys' collaborator, Desanto-Madeya. In Roy's community (sky blue) we observe a tendency to publish with different scholars, which is why her community does not represent a star topology (everybody is connected with everybody) in contrast to Fawcett's community. The star topology is also a characteristic of other communities in the network, such as the pink, red, green and light green clusters. The latter community is composed mainly of Spanish authors while the former ones are composed of authors from The United States.

**Figure 7:** Co-authorship network. Nodes represent authors, and links describe the number of publications co-authored. The size of the nodes is proportional to the degree of centrality of authors. Colors are assigned according to an algorithm of community detection. Data based on WoS.

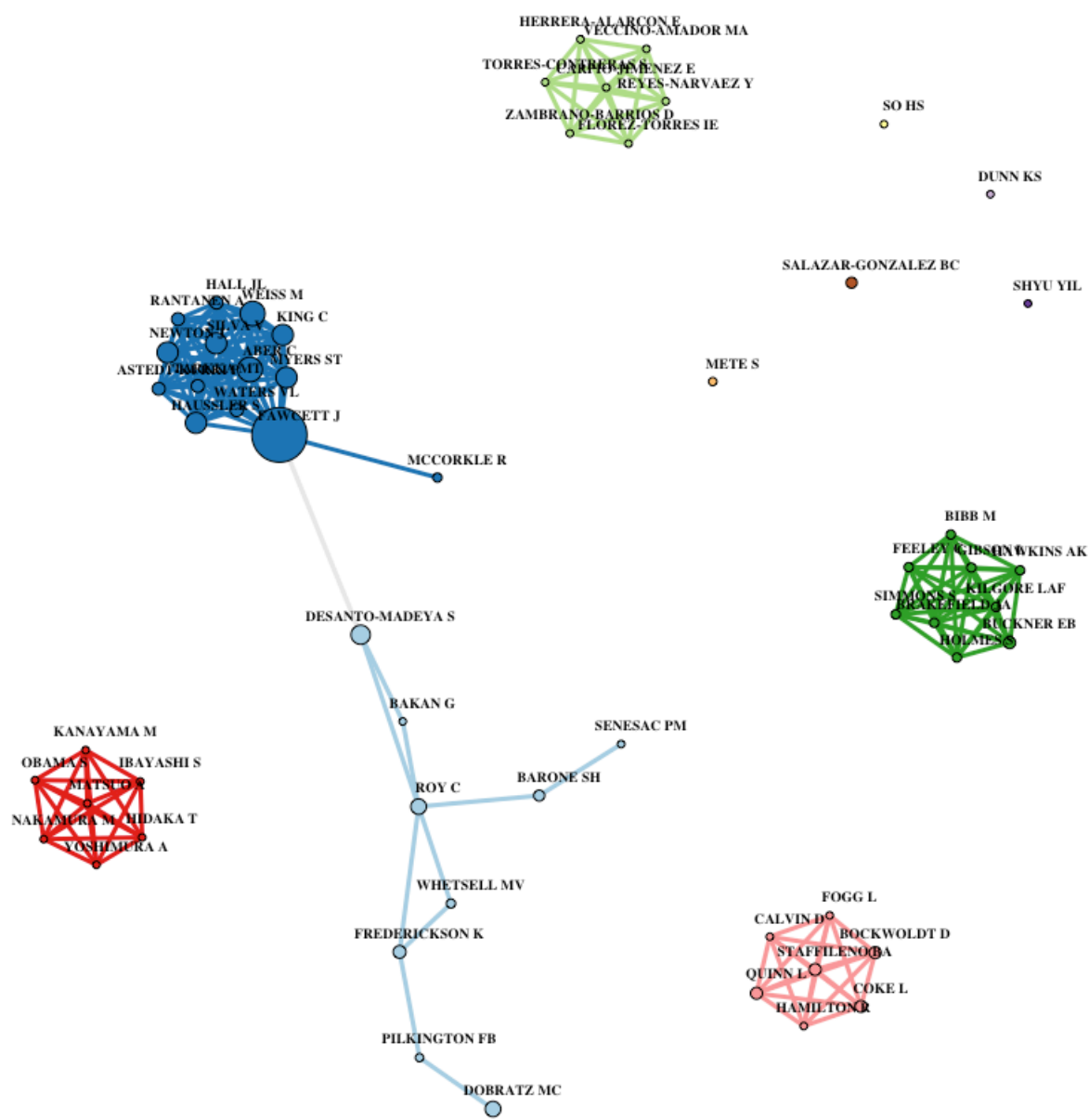

Regarding institutional collaboration, we map the co-authorship network at an institution level. Again, for illustrative purposes, in Figure 8 we present the main structure of the institutions network based on WoS (138 institutions), while in the Supplementary Material we attach a full institutions network based on SCOPUS (301 institutions). The network presented in Figure 8 shows that the largest *connected component* (a sub network in which there is a path between all the nodes) includes mainly US clusters, but one that is composed also of universities from China and Taiwan (red cluster). In this connected component we observe the blue cluster that is led by the University of Massachusetts Boston (Fawcetts' university) and the Boston College where Sor Callista Roy was affiliated for decades. Other clusters inside the same connected component are led by Marquette University (pink), University of Wisconsin (green), Yale University (red) and Lehman College with Oakland

University (sky blue). Other groups not linked to the largest connected component are for example those of Sweden (brown), Brazil (orange), Iran (light orange), or Spain (yellow), which stands out over other clusters because it includes two non-educational institutions linked to the University of Almería.

**Figure 8:** Institutional collaboration network. Nodes represent Institutions, and links show the number of publications co-authored. The size of the nodes is proportional to the degree of centrality of Institutions. Colors are assigned according to an algorithm of community detection. Data based on WoS.

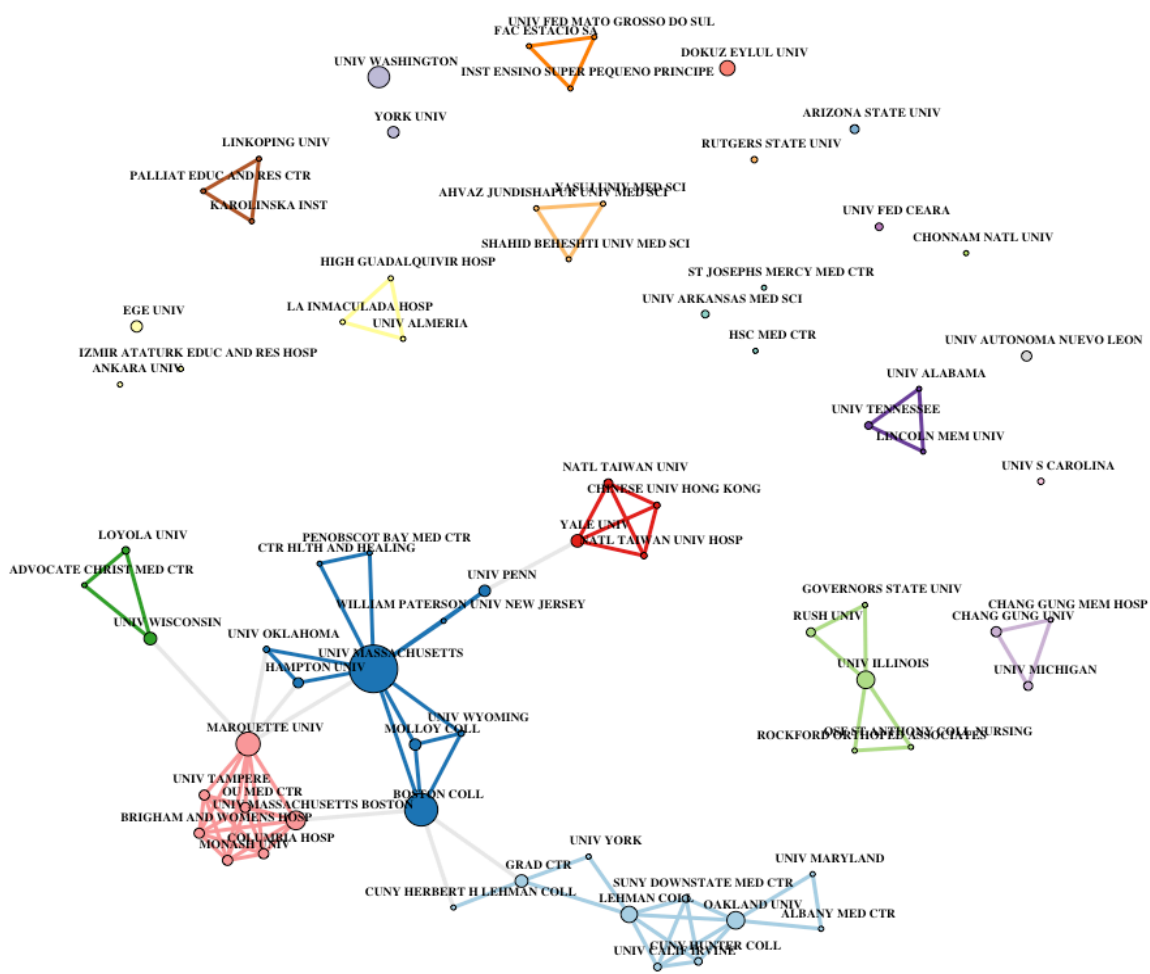

In a network we can measure quantitatively the level of centrality of each node, that is, how connected or how relevant a node is for the whole structure of the network. In this line, we have already mentioned degree centrality, which is the basic measure of centrality. We can add betweenness centrality and page rank to the measures of centrality. Betweenness centrality of a node, in this case of an institution, is the number of shortest paths that pass through the node. Page Rank is a measure of the influence of a node in the network (Page, Brin, Motwani, & Winograd, 1999). A high Page Rank score implies that an institution is connected to many other institutions who themselves have high scores. In Table 4, we show the ranking of each institution according to these three measures of centrality.

**Table 4**: Ranking of institutions according to three measures of centrality. Data based on WoS.

| Rank | Institution ID | Degree | Institution ID | Betweenness | Institution ID | Page Rank |
|------|----------------|--------|----------------|-------------|----------------|-----------|
| 1 | UNIV MASSACHUSETTS BOSTON | 0.073 | UNIV MASSACHUSETTS BOSTON | 0.036 | UNIV MASSACHUSETTS BOSTON | 0.0229 |
| 2 | MARQUETTE UNIV | 0.073 | BOSTON COLL | 0.035 | UNIV ILLINOIS | 0.0189 |
| 3 | BOSTON COLL | 0.051 | GRAD CTR | 0.027 | MARQUETTE UNIV | 0.0189 |
| 4 | UNIV MASSACHUSETTS | 0.051 | LEHMAN COLL | 0.023 | LEHMAN COLL | 0.0172 |

| | | | | | | BOSTON | |
| --- | --- | --- | --- | --- | --- | --- | --- |
| 5 | LEHMAN COLL | 0.051 | MARQUETTE UNIV | 0.022 | BOSTON COLL | 0.0165 |
| 6 | OAKLAND UNIV | 0.044 | UNIV PENN | 0.014 | OAKLAND UNIV | 0.0148 |
| 7 | OU MED CTR | 0.044 | UNIV WISCONSIN | 0.011 | EGE UNIV | 0.0136 |
| 8 | COLUMBIA HOSP | 0.044 | UNIV MASSACHUSETTS BOSTON | 0.011 | UNIV MICHIGAN | 0.0136 |
| 9 | UNIV TAMPERE | 0.044 | YALE UNIV | 0.011 | DOKUZ EYLUL UNIV | 0.0135 |
| 10 | MONASH UNIV | 0.044 | OAKLAND UNIV | 0.008 | UNIV AUTONOMA NUEVO LEON | 0.0135 |
| 11 | BRIGHAM AND WOMENS HOSP | 0.044 | HAMPTON UNIV | 0.004 | UNIV FED CEARA | 0.0135 |
| 12 | UNIV ILLINOIS | 0.037 | UNIV ILLINOIS | 0.001 | UNIV MASSACHUSETTS BOSTON | 0.0123 |
| 13 | YALE UNIV | 0.029 | EGE UNIV | 0.000 | UNIV WISCONSIN | 0.0123 |
| 14 | UNIV WISCONSIN | 0.029 | UNIV MICHIGAN | 0.000 | YALE UNIV | 0.0109 |
| 15 | GRAD CTR | 0.029 | DOKUZ EYLUL UNIV | 0.000 | OU MED CTR | 0.0104 |

As we have seen previously, the University of Massachusetts Boston confirms its leadership, ranking first in the three measures of centrality. Also, Boston College is second in betweenness centrality, which demonstrates that this institution is an important bridge between authors of different institutions. Marquete University is a highly connected institution, in second place (with the same score as that of first position) in terms of number of links or degree. Marquete University is also ranked in second position in the measure of influence Page Rank. In this measure, the University of Illinois and Lehman College are in the top five. Four universities outside of the United States appear in the list of the 15 most influential institutions (page rank column): the Turkish universities Dokuz Eylül and Egeo, the Mexican university Autónoma de Nuevo León and the Brazilian Federal University of Ceará. Regarding non-educational institutions, highly connected centers are the Oakland Medical Center, Columbia Hospital and Brigham and Women's Hospital.

# 5 Conclusions

The Roy's Adaptation Model has had an important impact on research in Nursing over more than forty years. We presented a global, longitudinal and descriptive analysis based on bibliometric indicators and data visualizations. We show an evolution of previous bibliometric studies by also applying analysis based on networks and using the state-of-the-art software of the field of bibliometrics.

The United States and its Institutions dominate research based on the Roy Adaptation Model. However, research topics in this field have been embraced by an extensive global network of scholars and institutions, including Latin American, European and Asian Universities. Articles comprise most of the research in the area, and an important number of them are published in the journal Nursing Science Quarterly.

The intellectual infrastructure of the Model is built mainly on books authored by Sor Callista Roy and on a few papers adapting the model. Dr. Fawcett and Sor Callista Roy play an important role in research conducted in the area, and similarly, University of Massachusetts Boston and Boston College stand out within the network of institutions that contribute to this research field. Non-educational Institutions such as the Oakland Medical Center or the Brighman and Women's Hospital have acquired an important ranking in the network of collaboration. Collaborative institutions with low levels of production do not have a central role in the network of institutions, and appear only in the full network provided as supplementary material.

Some countries of the chapters of the Roy Adaptive Association, such as Colombia or Panama, are still not presenting strong links with other countries and with other institutions. This situation may mean that in those countries, research based on the Roy Model is still conducted with colleagues from the same institution or published in articles that are still not indexed by WoS or SCOPUS. However, this is not the case of the University of Nuevo León in México that is considered among the 15 most influential institutions on the topic.

We think that our results are useful to Nursing Schools and also for experienced and novice researchers that would like to carry out research based on the Roy Adaptation Model, and who could search for topics of interest and for collaboration in a more informed way. Finally, our proposed analysis could be applied to other theoretical models in Nursing in a separate or joint manner.